\documentclass[aps,reprint,
twocolumn,prb,superscriptaddress,showpacs,preprintnumbers,
amsmath,amssymb,floatfix]{revtex4-1}
\pdfoutput=1

\usepackage{graphicx} 
\usepackage[english]{babel}
\usepackage[babel,kerning=true,spacing=true]{microtype}

\def\bG{{\bf G}}

\def\bk{{\bf k}}

\def\bp{{\bf p}}
\def\bq{{\bf q}}

\def\bG{{\bf G}}
\def\bQ{{\bf Q}}

\def\b0{{\bf 0}}

\def\cO{{\cal O}}

\def\bra{\langle}
\def\ket{\rangle}
\def\up{\uparrow}
\def\down{\downarrow}

\def\eps{\epsilon}

\def\om{\omega}

\def\sg{\sigma}

\def\sgn{{\rm sgn}}

\begin{document}

\title{Incommensurate nematic fluctuations in two dimensional metals}

\author{Tobias Holder}
\affiliation{Universit\"at Stuttgart, Fachbereich Physik,
 D-70550 Stuttgart, Germany}
\affiliation{Max-Planck-Institute for Solid State Research,
 D-70569 Stuttgart, Germany}
\author{Walter Metzner}
\affiliation{Max-Planck-Institute for Solid State Research,
 D-70569 Stuttgart, Germany}

\date{\today}

\begin{abstract}
To assess the strength of nematic fluctuations with a finite wave
vector in a two-dimensional metal, we compute the static $d$-wave 
polarization function for tight-binding electrons on a square 
lattice.
At Van Hove filling and zero temperature the function diverges 
logarithmically at $\bq = 0$.
Away from Van Hove filling the ground state polarization function 
exhibits finite peaks at finite wave vectors.
A nematic instability driven by a sufficiently strong attraction 
in the $d$-wave charge channel thus leads naturally to a 
spatially modulated nematic state, with a modulation vector that
increases in length with the distance from Van Hove filling.
Above Van Hove filling, for a Fermi surface crossing the magnetic 
Brillouin zone boundary, the modulation vector connects 
antiferromagnetic hot spots with collinear Fermi velocities.
\end{abstract}

\maketitle


\section{Introduction}

Metallic two-dimensional electron systems frequently exhibit 
multiple enhanced fluctuations in the charge, magnetic and pairing 
channels, which lead to a variety of competing instabilities.
In the last decade, nematic order and nematic fluctuations in 
two-dimensional metals have attracted increasing interest.
\cite{fradkin10}
In a homogeneous nematic state an orientational symmetry of the 
system is spontaneously broken, without breaking the translation 
invariance.
A nematic state can be formed by partial melting of stripe order 
in a doped antiferromagnetic Mott insulator.\cite{kivelson98}
Alternatively, a nematic state can be obtained from a Pomeranchuk
\cite{pomeranchuk59} instability generated by forward scattering
interactions in a normal metal.\cite{yamase00,halboth00}
On a square lattice, the most natural candidate for a Pomeranchuk
instability has a $d_{x^2-y^2}$ symmetry, which breaks the 
equivalence between $x$- and $y$-directions.

A direct experimental signature of nematic order is a pronounced
in-plane anisotropy in transport or spectroscopic measurements,
in the absence of additional Bragg peaks indicating a broken 
translation invariance.\cite{kivelson03}
Evidence for nematic order with a $d_{x^2-y^2}$ symmetry has
been observed in several strongly interacting electron compounds.
A nematic phase with a sharp phase boundary has been established 
in a series of experiments on ultrapure $\rm Sr_3 Ru_2 O_7$ 
crystals in a strong magnetic field.\cite{ruthenate}
Nematic order has also been observed in the high temperature
superconductor $\rm Y Ba_2 Cu_3 O_y$ in transport experiments
\cite{ando02,daou10} and neutron scattering.\cite{hinkov}
Due to the slight orthorhombicity of the $\rm CuO_2$ planes one 
cannot expect a sharp nematic phase transition in 
$\rm Y Ba_2 Cu_3 O_y$. However, the strong temperature dependence
of the observed in-plane anisotropy indicates that the system 
develops an intrinsic electronic nematicity, which enhances the 
in-plane anisotropy imposed by the structure.
\cite{yamase06,yamase09}

Nematic fluctuations close to a continuous nematic quantum phase
transition induce non-Fermi liquid behavior with a strongly 
momentum dependent decay rate of electronic excitations.
\cite{oganesyan01,metzner03,dellanna06,dellanna07}
Mean-field theories for nematic transitions driven by forward
scattering interactions in two dimensions indicate that the
transition is typically first order at low temperatures,
\cite{kee03,khavkine04,yamase05} such that a quantum critical
point remains elusive.
However, order parameter fluctuations can change the order of
the transition, in favor of a continuous transition even at 
zero temperature.\cite{jakubczyk09,yamase11}

So far, only homogeneous nematic states have been considered.
However, Metlitski and Sachdev \cite{metlitski10a,metlitski10b}
recently found a tendency toward formation of a {\em modulated} 
nematic state in a two-dimensional metal with strong
antiferromagnetic spin-density wave fluctuations.
In that state the nematic order oscillates spatially across the
crystal, with a small and generally incommensurate wave vector 
that points along the Brillouin zone diagonal and increases in
length with the distance of the Fermi surface from the Van Hove 
points.
The modulated nematic order occurs as a secondary instability 
generated by the antiferromagnetic spin fluctuations, and is
related by an emergent pseudospin symmetry to the familiar
$d$-wave pairing instability.\cite{metlitski10b}

In this paper, we investigate the possibility of a modulated
nematic state from a different starting point.
We consider a model of tight-binding electrons on a square
lattice with an interaction that has an attractive $d$-wave 
component for forward scattering in the charge channel.
Within a random phase approximation (RPA), we analyze at which 
wave vector the modulated nematic fluctuations are maximal.
For electron densities above Van Hove filling we find the 
same diagonal wave vector as Metlitski and Sachdev in their 
spin-density wave model.\cite{metlitski10a,metlitski10b}
Below Van Hove filling a modulation parallel to the $x$- or
$y$-axis can be favored.

In realistic models nematic fluctuations compete with other 
potential instabilities such as magnetism and superconductivity.
Here we do not address this competition. Our model interaction 
is restricted to a specific $d$-wave charge channel.
The competition of modulated nematic fluctuations with
antiferromagnetism, superconductivity and charge density
wave order in the two-dimensional (extended) Hubbard model 
has been the subject of a recent renormalization group
analysis.\cite{husemann12}

The paper is structured as follows.
In Sec.~II we define our model and introduce the $d$-wave 
charge susceptibility.
In Sec.~III we analyze the momentum dependence of the static 
$d$-wave charge susceptibility. We identify the peaks which
determine the wave vector of the leading instability in the
ground state, and we compute the singular momentum dependence
around the peaks.
We also determine the shift of the peaks at low finite 
temperature.
We finally conclude in Sec.~IV.


\section{Model and susceptibility}

We consider a one-band model of electrons on a square lattice
with a dispersion $\eps_{\bk}$ and an interaction of the form 
\cite{metzner03}
\begin{equation} \label{H_I}
 H_I = 
 \frac{1}{2L} \sum_{\bq} g(\bq) \, n_d(\bq) \, n_d(-\bq) \; ,
\end{equation}
where $L$ is the number of lattice sites, and
\begin{equation} \label{n_d}
 n_d(\bq) = \sum_{\bk} \sum_{\sg=\up,\down} d_{\bk} \,
 c_{\bk-\bq/2,\sg}^{\dag} c_{\bk+\bq/2,\sg}
\end{equation}
is a $d$-wave density fluctuation operator;
$d_{\bk}$ is a form factor with $d_{x^2-y^2}$ symmetry such as
$d_{\bk} = \cos k_x - \cos k_y$.
The coupling function $g(\bq)$ is negative and may depend on
the momentum transfer $\bq$. 
An interaction of the form $H_I$ appears in the $d$-wave charge 
channel of the two-particle vertex in microscopic models such 
as the Hubbard or $t$-$J$ model.\cite{yamase00,halboth00}
Note that the fermionic operators in Eq.~(\ref{n_d}) are taken
at two momenta $\bk \pm \bq/2$ situated symmetrically around 
the momentum $\bk$ appearing in the form factor $d_{\bk}$.
Hence, for $\bq = \bQ = (\pi,\pi)$, the operator $n_d(\bq)$ 
differs significantly from the socalled $d$-density wave 
order parameter
$\sum_{\bk,\sg} d_{\bk} c_{\bk + \bQ,\sg}^{\dag} c_{\bk,\sg}$.
\cite{chakravarty01}

For the kinetic energy we use a tight-binding dispersion of
the form
\begin{eqnarray} \label{eps_k}
 \eps_{\bk} &=& -2t (\cos k_x + \cos k_y) - 4t' \cos k_x \cos k_y\nonumber\\
 &&- 2t'' (\cos 2k_x + \cos 2k_y) \; ,
\end{eqnarray}
where $t$, $t'$, and $t''$ are the hopping amplitudes between
nearest, next-to-nearest, and third-nearest neighbors on the
square lattice, respectively.
We assume $t > 0$, $t' < 0$ and $t'' \geq 0$ as is adequate 
for cuprate superconductors.
The dispersion has saddle points at $(\pi,0)$ and $(0,\pi)$
for $t'' \leq {t''}^* = (t + 2t')/4$. 
For $t'' \geq {t''}^*$ the saddle points are shifted to
$\pm \big(\pi - \arccos\frac{{t''}^*}{t''},0 \big)$ and
$\pm \big(0,\pi - \arccos\frac{{t''}^*}{t''} \big)$.
A special situation (Van Hove filling) arises when the Fermi
surface touches the saddle points, so that the density of
states diverges logarithmically at the Fermi level. 
The chemical potential corresponding to Van Hove filling
is given by
\begin{equation} \label{mu_vh}
 \mu_{\rm vh} = \left\{ \begin{array}{cc}
 4t' - 4t'' & 
 \quad \mbox{for} \quad t'' \leq {t''}^* \\
 \left( \frac{1}{4} t^2 + tt' + {t'}^2 - 2tt'' \right)/t'' &
 \quad \mbox{for} \quad t'' \geq {t''}^*
 \end{array} \right. \; .
\end{equation}

The amount of $d$-wave charge fluctuations can be quantified
by the dynamical $d$-wave charge (density) susceptibility
\cite{dellanna06,yamase04}
\begin{equation} \label{N_d}
 N_d(\bq,\om) = - i \int_0^{\infty} dt \, e^{i\om t} \,
 \bra [ n_d(\bq,t), n_d(-\bq,0) ] \ket \; ,
\end{equation}
where $n_d(\bq,t)$ is the time dependent operator corresponding 
to $n_d(\bq)$ in the Heisenberg picture.
Within RPA, the $d$-wave charge susceptibility in the model
defined above is given by
\begin{equation} \label{RPA}
 N_d(\bq,\om) = 
 \frac{2\Pi_d^0(\bq,\om)}{1 - 2 g(\bq) \Pi_d^0(\bq,\om)} \; ,
\end{equation}
with the bare $d$-wave polarization function (particle-hole 
bubble)
\begin{equation} \label{Pi_d^0}
 \Pi_d^0(\bq,\om) = - \int \frac{d^2p}{(2\pi)^2} \,
 \frac{f(\xi_{\bp+\bq/2}) - f(\xi_{\bp-\bq/2})}
 {\om + i0^+ - (\eps_{\bp+\bq/2} - \eps_{\bp-\bq/2})} \,
 d_{\bp}^2  \; .
\end{equation}
Here and in the following $f$ is the Fermi function and
$\xi_{\bk} = \eps_{\bk} - \mu$.
The factor 2 in Eq.~(\ref{RPA}) is due to the spin sum.

An instability toward an ordered state with an order parameter
$\bra n_d(\bq^*) \ket \neq 0$ is signalled by a divergence of 
the static susceptibility $N_d(\bq,0)$ for a certain wave vector 
$\bq^*$. Within RPA the instability is reached once 
$2g(\bq^*) \Pi_d^0(\bq^*,0) = 1$ while $2g(\bq) \Pi_d^0(\bq,0) < 1$
for $\bq \neq \bq^*$.
For $\bq^* = \b0$ the ordered state is a homogeneous nematic 
with unbroken translation invariance. The case $\bq^* \neq \b0$
leads to a modulated nematic state with a modulation vector
$\bq^*$.
In previous studies of the above model it was always assumed 
that the coupling function $g(\bq)$ is sufficiently strongly 
peaked at $\bq = \b0$ such that the leading instability is at 
$\bq^* = \b0$.
In the present paper we consider the case where $g(\bq)$ 
exhibits no or only a weak dependence on $\bq$ in some region
around $\bq = \b0$.
The leading instability in the $d$-wave charge channel then 
occurs at wave vectors at which $\Pi_d^0(\bq,0)$ has a peak.
In the following we therefore study the structure of the
$d$-wave particle-hole bubble, paying particular attention to
its extrema.


\section{Static particle-hole bubble}

In this section we analyze the momentum dependence of the
$d$-wave particle-hole bubble $\Pi_d^0$ at zero frequency.
We will also consider the usual ($s$-wave) particle-hole
bubble $\Pi^0$ for comparison. The latter is given by 
Eq.~(\ref{Pi_d^0}) without the $d$-wave form factor.
The structure depends significantly on the electron density
which is determined by the chemical potential.
There are three qualitatively different cases: below, at, 
and above Van Hove filling.
We will first analyze $\Pi_d^0(\bq,0)$ in the ground state, 
and treat finite temperatures subsequently.

\subsection{Global structure and peaks}
In Fig.~1 we show $\Pi_d^0(\bq,0)$ in the ground state as 
a function of $\bq$, for hopping parameters $t = 1$, 
$t' = - 1/4$, and $t'' = 0$, with three different choices 
for the electron density below, at, and above Van Hove filling,
corresponding to $\mu = \mu_{\rm vh} - 0.01$, $\mu = \mu_{\rm vh}$,
and $\mu = \mu_{\rm vh} + 0.05$, respectively.
In all cases there are enhanced (negative) values along 
the lines in the Brillouin zone given by the condition
\begin{figure}[t]
\begin{center}
\includegraphics[width=3.2in]{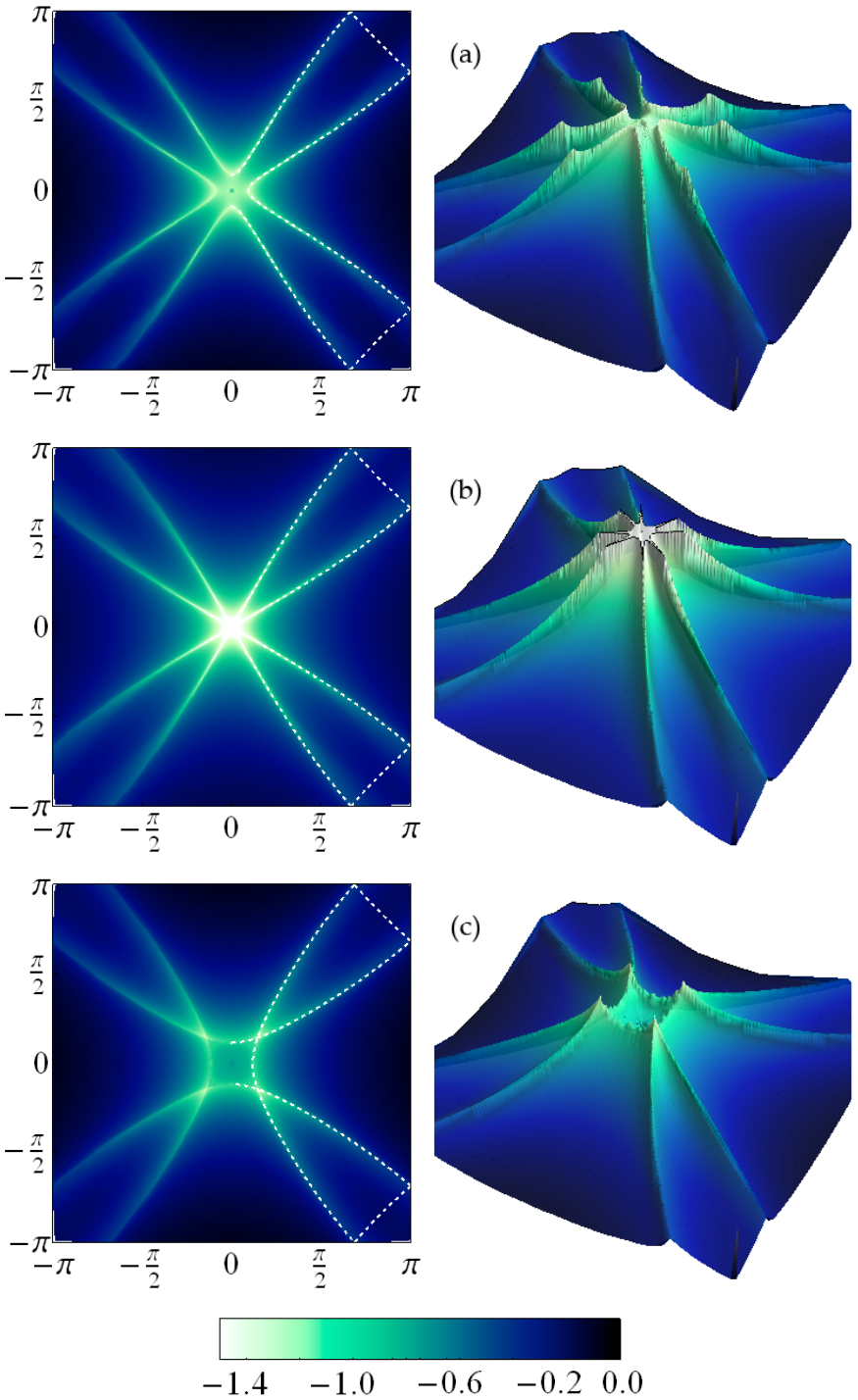}
\caption{(Color online) $d$-wave particle-hole bubble 
 $\Pi_d^0(\bq,0)$ as a function of $\bq$ at temperature
 zero for an electron density below (a), at (b), and 
 above (c) Van Hove filling. 
 The plots on the right are 3D and sign-reversed versions 
 of the color-coded planar intensity plots on the left.
 In the 3D plot at Van Hove filling, the divergence at
 the origin has been cut off.
 The dashed lines drawn in the planar plots are defined 
 by the condition $\xi_{(\bq + \bG)/2} = 0$.}
\end{center}
\end{figure}
\begin{equation} \label{lines}
 \xi_{(\bq + \bG)/2} = 0 \; ,
\end{equation}
where $\bG$ is a reciprocal lattice vector.
Geometrically these lines can be constructed by expanding
the Fermi surface by a factor two and then backfolding
pieces outside the first Brillouin zone into the first 
zone, as illustrated in Fig.~2. 
The momentum transfers $\bq$ satisfying the condition 
(\ref{lines}) are special in that they connect Fermi 
points with parallel tangents.
Eq.~(\ref{lines}) is the lattice generalization of the 
condition $|\bq| = 2k_F$ in a continuum system with a 
circular Fermi surface. With this in mind, one may call
such momentum transfers {\em $2k_F$-momenta} on the 
lattice, too, although the Fermi momenta in a lattice 
system do not have a common modulus $k_F$.\cite{altshuler95}
In the following we will refer to the lines defined by 
Eq.~(\ref{lines}) as {\em $2k_F$-lines}.
\begin{figure}[t]
\begin{center}
\includegraphics[width=2.5in]{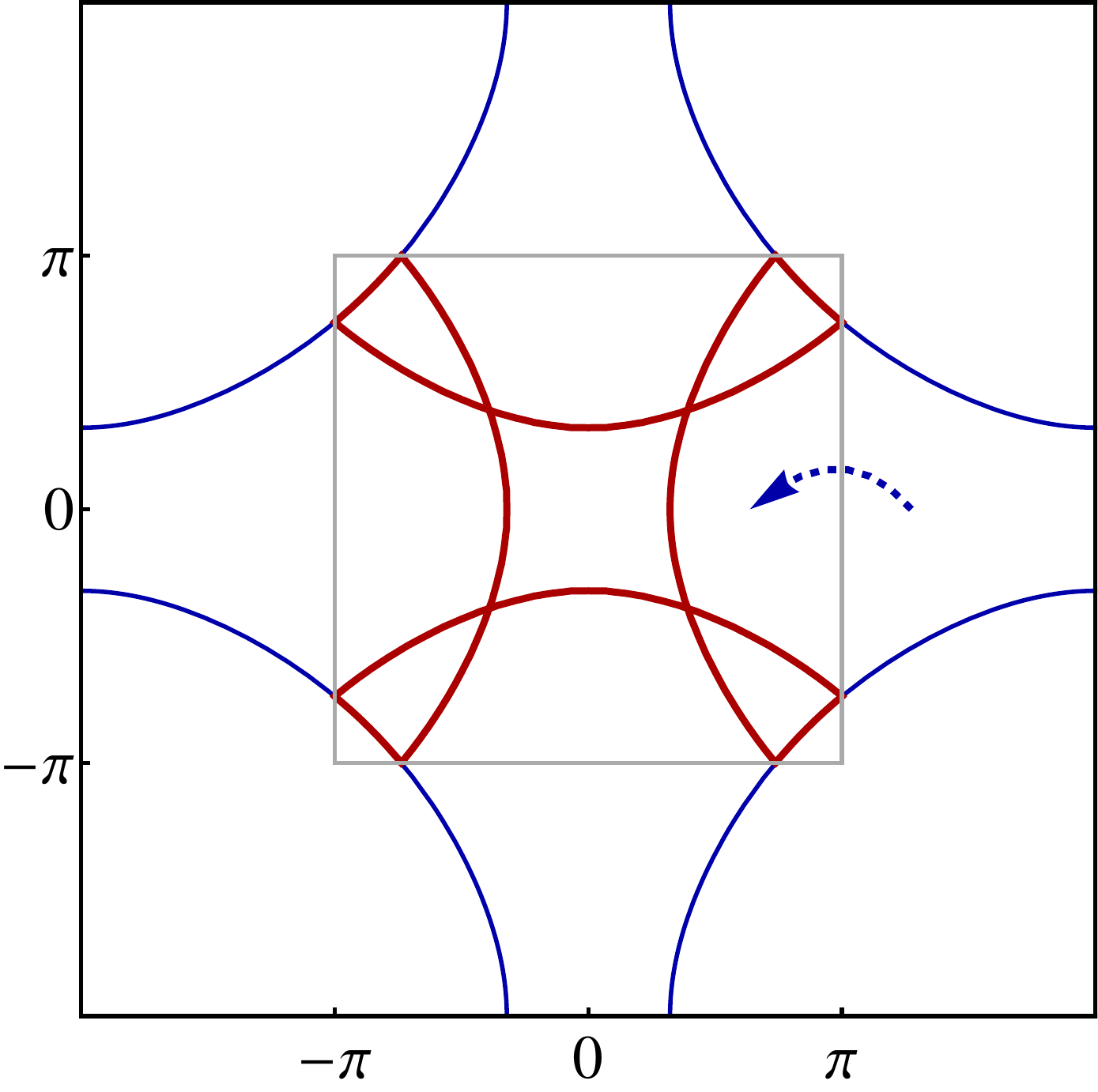}
\caption{(Color online) Construction of the $2k_F$-lines
 by backfolding the expanded Fermi surface into the first
 Brillouin zone.}
\end{center}
\end{figure}

In Fig.~3 we show the $s$-wave particle-hole bubble 
$\Pi^0(\bq,0)$ for comparison. 
There is again some structure tracing the lines given by 
$\xi_{(\bq + \bG)/2}$. 
However, the strongest weight now lies in a region near 
$(\pi,\pi)$ in the Brillouin zone, which is generated by
particle-hole excitations connecting the saddle point
regions near $(\pi,0)$ and $(0,\pi)$.
In the $d$-wave bubble these contributions are suppressed
by the $d$-wave form factor, since $d_{\bp} = 0$ for 
$\bp = (\pm\frac{\pi}{2},\pm\frac{\pi}{2})$.
Note that this suppression hinges upon the symmetric choice 
of the momentum in $d_{\bp}$ at the center of the fermionic
momenta $\bp \pm \bq/2$.
To obtain the polarization bubble describing $d$-density wave
fluctuations as defined by Chakravarty et al.\cite{chakravarty01}
one would have to replace $\bp - \bq/2$ by $\bp$ and 
$\bp + \bq/2$ by $\bp + \bq$ in Eq.~(\ref{Pi_d^0}).
For $\bq = \bQ = (\pi,\pi)$ one then picks up large
contributions from the saddle point region.

Incommensurate charge density waves with conventional
$s$-wave symmetry have also been discussed for two-dimensional
electron systems, especially in the context of cuprate 
superconductors.\cite{castellani95}
In that scenario, the incommensurate modulation vector is
however determined by a competition between Coulomb energy
and phase separation tendencies, not by peaks in the
polarization function $\Pi^0(\bq,0)$.
Peaks in $\Pi^0(\bq,0)$ at small incommensurate wave vectors 
occurring for large doping (far from half-filling) can be 
associated with modulated {\em ferromagnetic} fluctuations, 
for example, in a two-dimensional Hubbard model.
\cite{igoshev10,igoshev11}

At Van Hove filling, the $d$-wave particle-hole bubble is
strongly peaked at $\bq = \b0$, which is a crossing point 
of eight lines satisfying $\xi_{(\bq + \bG)/2} = 0$.
The leading $d$-wave charge instability is therefore a 
homogeneous nematic one in this case.

\begin{figure}[t]
\begin{center}
\includegraphics[width=3.2in]{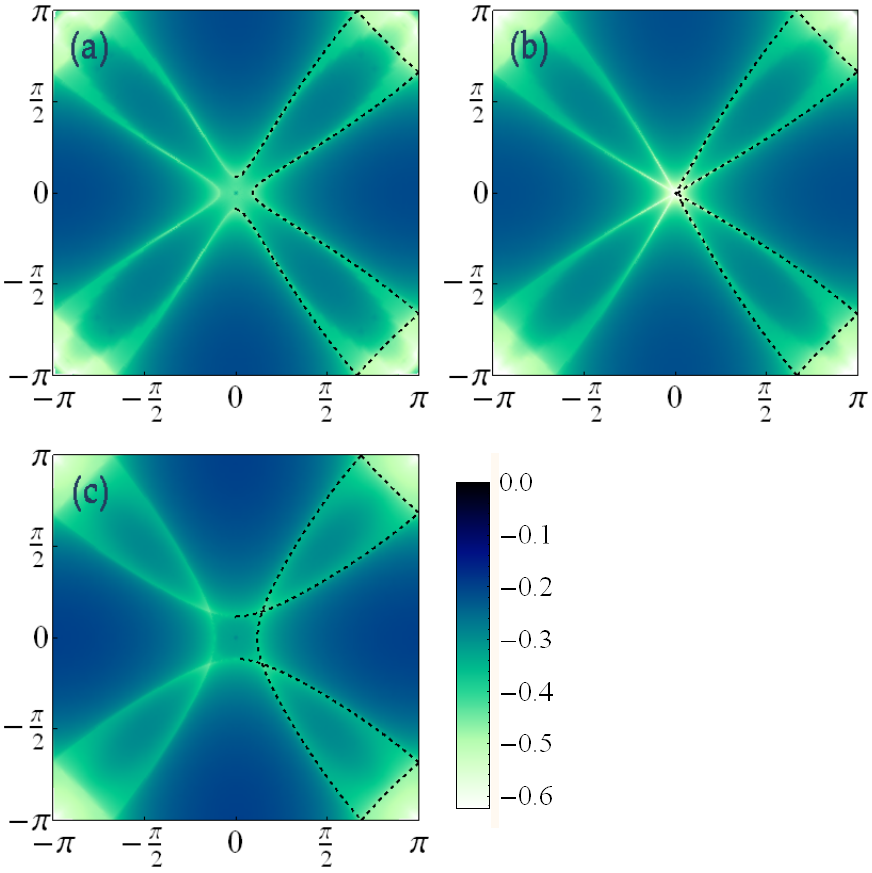}
\caption{(Color online) $s$-wave particle-hole bubble 
 $\Pi^0(\bq,0)$ as a function of $\bq$ for an electron
 density below (a), at (b), and above (c) Van Hove filling. 
 The dashed lines are defined by the condition 
 $\xi_{(\bq + \bG)/2} = 0$.}
\end{center}
\end{figure}

Below but close to Van Hove filling, the highest weight is 
obtained at those points in $\bq$-space where the lines given 
by Eq.~(\ref{lines}) are closest to the origin, that is, 
on the $q_x$- and $q_y$-axes.
The leading instability is thus a modulated nematic with
a small modulation vector along one of the crystal axes, 
or a superposition of such modulations.
Let us denote the distance of the points of highest 
weight from the origin by $q_a$.
Solving $\xi_{(\bq + \bG)/2} = 0$ for $\bq = (q_a,0)$ 
and $\bG = - (2\pi,0)$, one obtains
\begin{eqnarray} \label{q_a}
 q_a &=& 2 \arccos \left( 
 \frac{t + 2t' - \sqrt{(t + 2t')^2 - 4t''(2t + \mu)}}{4t''}
 \right)
 \nonumber \\ 
 &=& 2 \arccos \left( \frac{2t + \mu}{2t + 4t'} \right)
 \quad \mbox{for} \quad t'' = 0
 \; .
\end{eqnarray}
Further away from Van Hove filling, the global maxima of
$|\Pi_d^0(\bq,0)|$ can be situated at inflection points of
the $2k_F$-lines, which are inherited from inflection points
of the Fermi surface. Explicit expressions for these points
are rather lengthy so that we refrain from reporting them.
They are not linked to any symmetry axis of the lattice.

Above Van Hove filling, the distance of the lines 
Eq.~(\ref{lines}) from the origin is given by 
\begin{eqnarray} \label{q'_a}
 q_a &=& 2 \arccos \left( 
 \frac{-t + 2t' + \sqrt{(t - 2t')^2 - 4t''(2t - \mu)}}{4t''}
 \right)
 \nonumber \\ 
 &=& 2 \arccos \left( \frac{2t - \mu}{2t - 4t'} \right)
 \quad \mbox{for} \quad t'' = 0
 \; .
\end{eqnarray}
However, the points $(q_a,0)$ etc.\ are not the global 
extrema.
Above Van Hove filling, the lines Eq.~(\ref{lines}) 
intersect on the diagonals in the Brillouin zone, and
the heighest weight is reached at these intersection 
points.
The leading instability is thus a modulated nematic with
a diagonal modulation vector $\bq^* = (\pm q_d, \pm q_d)$, 
or a superposition of such modulations.
Solving $\xi_{(\bq + \bG)/2} = 0$ for $\bq = (q_d,q_d)$ 
and $\bG = - (2\pi,0)$ or $\bG = - (0,2\pi)$, one obtains
\begin{eqnarray} \label{q_d}
 q_d &=& 2 \arccos \sqrt{\frac{\mu - 4t''}{4t' - 8t''}}
 \nonumber \\ 
 &=& 2 \arccos \sqrt{\frac{\mu}{4t'}}
 \quad \mbox{for} \quad t'' = 0
 \; .
\end{eqnarray}
Close to Van Hove filling, $q_d$ is small. Note that $q_d$
is defined only for $\mu_{\rm vh} \leq \mu \leq 4t''$. 
For $\mu > 4t''$, no intersection points exist.

\begin{figure}[b]
\begin{center}
\includegraphics[width=2.5in]{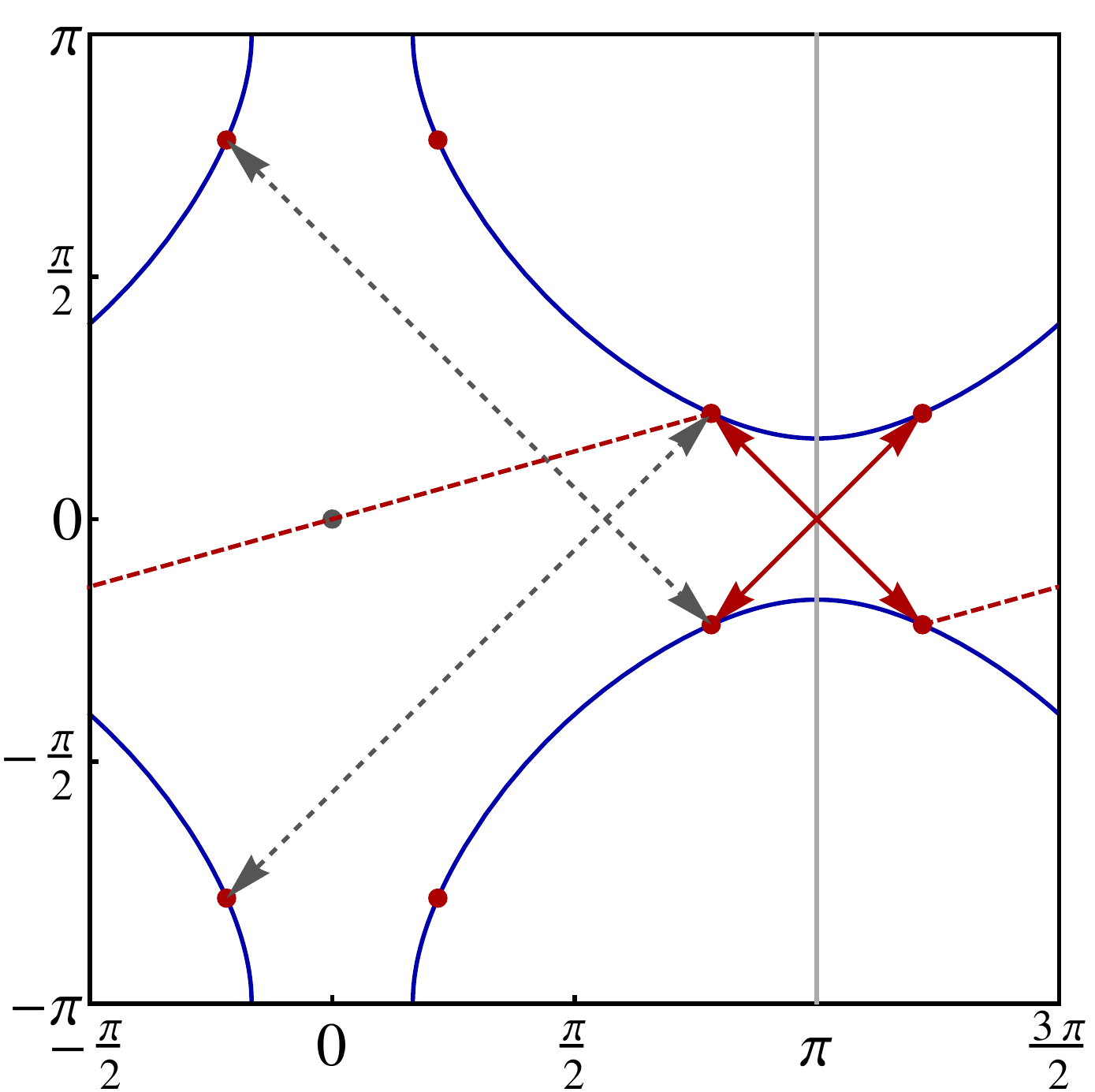}
\caption{(Color online) Hot spots and special transfer momenta. 
 The red solid lines with arrows represent the incommensurate 
 modulation vectors $(\pm q_d, \pm q_d)$, the grey dotted lines
 represent momenta $(\pm\pi,\pm\pi)$.
 The momentum represented by the long dashed red line can be
 reduced to one of the short incommensurate momenta by
 adding a reciprocal lattice vector.}
\end{center}
\end{figure}

The wave vector $\bq_d = (q_d,q_d)$ connects antiferromagnetic
{\em hot spots} on the Fermi surface (see Fig.~4).
This is because these hot spots lie on the magnetic Brillouin
zone boundary and have parallel Fermi surface tangents, and are 
therefore connected by a diagonal ``$2k_F$-vector''.
The length of $\bq_d$ is thus twice the distance between the 
hot spots and the nearest $M$-point $(\pi,0)$ or $(0,\pi)$.
An instability toward a modulated nematic state with the
same modulation vector $\bq_d$ has been obtained in a recent 
study of secondary instabilities generated by antiferromagnetic 
fluctuations.\cite{metlitski10a,metlitski10b}
In that analysis, $\bq_d$ arises naturally as a vector connecting 
hot spots since these are the points where (commensurate) 
antiferromagnetic fluctuations couple most strongly to electronic 
excitations at the Fermi surface. 
It is remarkable that the same modulation vector emerges
as a peak in the $d$-wave particle-hole bubble, where it
is selected by the Fermi surface geometry without any 
relation to antiferromagnetism.

\subsection{Expansion around $\bq = \b0$}

Away from Van Hove filling, there is a region around
$\bq = \b0$ where $\Pi_d^0(\bq,0)$ is relatively flat and
isotropic (see Fig.~1). This region is clearly limited 
by the $2k_F$-lines $\xi_{(\bq + \bG)/2} = 0$.
For $|\bq| \ll q_a$ one may expand around $\bq = 0$.
The leading terms are \cite{dellanna06}
\begin{equation} \label{small_q}
 \Pi_d^0(\bq,0) = a + c |\bq|^2 + \cO(|\bq|^4)
 \; .
\end{equation}
The first coefficient is given by a weighted density
of states at the Fermi level,
\begin{equation} \label{a}
 a = - N_{d^2}(\mu) = - \int \frac{d^2k}{(2\pi)^2} \,
 d_{\bk}^2 \delta(\mu - \eps_{\bk}) \; ,
\end{equation}
with a minus sign, so that $a$ is always negative.
The second coefficient can be written in the form
\cite{dellanna06}
\begin{equation} \label{c}
 c = \frac{1}{16} N'_{d^2\Delta\eps}(\mu) -
 \frac{1}{48} N''_{d^2v^2}(\mu) \; .
\end{equation}
Here
\begin{equation} 
 N_{d^2\Delta\eps}(\eps) = \int \frac{d^2k}{(2\pi)^2} \, 
 d_{\bk}^2 \Delta\eps_{\bk} \, \delta(\eps - \eps_{\bk})
 \; ,
\end{equation}
with $\Delta = \partial_{k_x}^2 + \partial_{k_y}^2$, and
\begin{equation} 
 N_{d^2 v^2}(\eps) = \int \frac{d^2k}{(2\pi)^2} \, 
 d_{\bk}^2 v_{\bk}^2 \, \delta(\eps - \eps_{\bk})
 \; ,
\end{equation}
with $v_{\bk} = |\nabla\eps_{\bk}|$, and the primes denote
derivatives.
Near Van Hove filling, the coefficient $c$ is dominated by
the first term in Eq.~(\ref{c}), since 
$N_{d^2\Delta\eps}(\mu)$ diverges for $\mu \to \mu_{\rm vh}$,
while $N_{d^2 v^2}(\mu)$ remains finite. The sign of $c$ 
is typically positive for $\mu < \mu_{\rm vh}$, and negative
for $\mu > \mu_{\rm vh}$.

Hence, away from Van Hove filling, the interaction term $H_I$ 
from Eq.~(\ref{H_I}) generates a homogeneous nematic instability 
only if $g(\bq)$ decays sufficiently rapidly at finite $\bq$.
That is, below Van Hove filling, $g(\bq_a,0) \Pi_d^0(\bq_a,0)$ with
$\bq_a = (q_a,0)$ has to be smaller than $g(\b0) \Pi_d^0(\b0,0)$.
Furthermore, above Van Hove filling, 
$g(\bq_d,0) \Pi_d^0(\bq_d,0)$ with $\bq_d = (q_d,q_d)$ needs 
to be smaller than $g(\b0) \Pi_d^0(\b0,0)$, and the curvature
of $g(\bq)$ around the origin has to compensate for the 
negative curvature of $\Pi_d^0(\bq,0)$.
The RPA analysis therefore indicates that an instability
toward a modulated nematic state is more natural.
On the other hand, fluctuations beyond RPA may smear out the
peaks resulting from the Fermi surface geometry, which could
favor a homogeneous nematic state over a modulated one. 
In principle, such fluctuations may also wipe out a nematic 
instability completely.\cite{yamase11}

\subsection{Shape of ridges and expansion around $\bq^*$}

The maxima of $|\Pi_d^0(\bq,0)|$ lie on ridges following the 
$2k_F$-lines given by $\xi_{(\bq+\bG)/2} = 0$ in the Brillouin 
zone.
The height of the ridges evolves regularly along these lines,
with few exceptions.
The shape of the ridge determined by the $\bq$-dependence of 
$\Pi_d^0(\bq,0)$ perpendicular to the lines is generically 
singular.
To discuss the form of this singularity, we introduce a 
coordinate $q_r$ describing the oriented distance from the 
closest $2k_F$-line, see Fig.~5.
We denote points on the $2k_F$-line by $\bq_{2k_F}$, and the 
radius of curvature at $\bq_{2k_F}$ by $2m_F v_F$, where $v_F$ 
is the electron velocity at the corresponding point $\bk_F$ on 
the Fermi surface (the radius of curvature of the Fermi surface 
at $\bk_F$ is $m_F v_F$).
We consider the generic case where the curvature at $\bq_{2k_F}$ 
is finite. Exceptional cases with vanishing curvature exist at
Van Hove points and at inflection points of non-convex Fermi 
surfaces.
The coordinate $q_r$ is defined positively on the ``outer'' side
of the $2k_F$-line, that is, the side containing the tangent to 
the line at $\bq_{2k_F}$, and negatively on the ``inner'' side.
\begin{figure}[ht!]
\begin{center}
\includegraphics[width=1.0in]{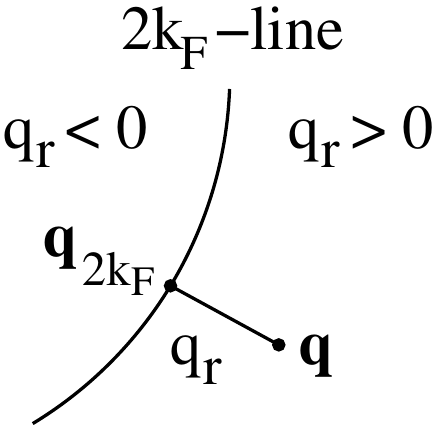}
\caption{$2k_F$-line and coordinate $q_r$.}
\end{center}
\end{figure}

Before describing the $2k_F$-singularity of $\Pi_d^0(\bq,0)$, 
it is instructive to recall the behavior of the $s$-wave bubble
$\Pi^0(\bq,0)$ for a quadratic dispersion $\eps_{\bk} = |\bk|^2/2m$,
which can be expressed in term of elementary functions:
\cite{stern67}
\begin{equation} \label{stern}
 \Pi^0(\bq,0) = - \frac{m}{2\pi} +
 \Theta(|\bq| - 2k_F) \frac{m k_F}{\pi|\bq|}
 \sqrt{\left( \frac{|\bq|}{2k_F} \right)^2 - 1}
 \; .
\end{equation}
Note that $\Pi^0(\bq,0)$ is constant for all momenta satisfying
$|\bq| \leq 2k_F$.
At the $2k_F$-line, here given by $|\bq| = 2k_F$, the $s$-wave 
bubble has a square-root singularity with infinite slope on the 
outer side.
For small $q_r = |\bq| - 2k_F$, the bubble has the form
\begin{equation} \label{stern_exp}
 \Pi^0(\bq,0) = - \frac{m}{2\pi} +
 \Theta(q_r) \frac{\sqrt{m q_r/v_F}}{2\pi} + \cO(q_r^{3/2}) \; .
\end{equation}

The momentum dependence of the $d$-wave particle-hole bubble 
near a $2k_F$-line for two-dimensional lattice electrons
has a similar form,
\begin{equation} \label{near2k_F}
 \Pi_d^0(\bq,0) =  \Pi_d^0(\bq_{2k_F},0) +
 \Theta(q_r) \frac{d_{\bk_F}^2}{2\pi} \sqrt{m_F q_r/v_F} + 
 b_F q_r \; ,
\end{equation}
where $\bk_F$ is the Fermi momentum corresponding to the
point $\bq_{2k_F}$ on the $2k_F$-line, $v_F$ and $m_F v_F$
are the the Fermi velocity and the Fermi surface curvature
at $\bk_F$, respectively, and $b_F$ is another constant.
The square-root singularity in this expression is essentially
the same as in Eq.~(\ref{stern_exp}).
The absence of a term linear in $q_r$ in Eq.~(\ref{stern_exp})
is a pecularity of $\Pi^0$ for a quadratic dispersion relation
in two dimensions. Usually $b_F$ is a negative number, such
that $|\Pi_d^0(\bq,0)|$ decreases with increasing $|q_r|$ in
both directions.

The momentum dependence perpendicular to the ridge in 
Eq.~(\ref{near2k_F}) is closely analogous to the corresponding
behavior of the s-wave bubble $\Pi^0$ in a two-dimensional
lattice system,\cite{altshuler95} the only difference being
the factor $d_{\bk_F}^2$ for the d-wave case.
Altshuler et al.\cite{altshuler95} also specified the momentum
dependence for small tangential shifts, in analogy to the
isotropic case. That dependence, however, is generally 
modified by a non-universal variation of the bubble along the
ridge.

At points of zero curvature, in particular inflection points,
$m_F$ diverges and the expression (\ref{near2k_F}) is not
applicable. For $t'/t < 0$ such points exist typically below
Van Hove filling, and they can host the global extrema of
$\Pi_d^0(\bq,0)$.
Above Van Hove filling the global extrema are situated at
the crossing points $\bq^* = (\pm q_d, \pm q_d)$ of two 
$2k_F$-lines on the Brillouin zone diagonal (see Sec.\ III.A). 
The momentum dependence of $\Pi_d^0(\bq,0)$ near $\bq^*$ is
then given by a superposition of two ridges of the form
(\ref{near2k_F}), which are mirror symmetric with respect
to the diagonal. The momentum dependence is linear in the
edge bounded by the inner side of both ridges, while it is
dominated by a square-root singularity with infinite slope
in the other three edges formed by the crossing ridges.

\subsection{Finite temperature}

At finite temperatures the singularities of the particle-hole 
bubble are smoothed and the peaks are generally shifted with 
respect to their ground state position. 
In this section we analyze these effects at low finite $T$.

The particle-hole bubble at $T>0$ can be written as a 
convolution of the bubble at $T=0$ with the energy-derivative
of the Fermi function:
\begin{equation} \label{convol}
 \Pi_d^0(\bq;T,\mu) = \int_{-\infty}^{\infty} d\mu' \,
 h(\mu - \mu') \, \Pi_d^0(\bq;0,\mu') \; ,
\end{equation}
where
\begin{equation}
 h(\xi) = -f'(\xi) = 
 \frac{1}{4T \cosh^2\left(\frac{\xi}{2T} \right)} \, .
\end{equation}
Note that we have suppressed the frequency variable ($\om = 0$) 
in the argument of $\Pi_d^0$ while making the dependences on 
$T$ and $\mu$ explicit. Note also that $\int d\xi \, h(\xi) = 1$.

We now determine how the ridges along the $2k_F$-lines are
shifted and smoothed at $T>0$.
To this end, we parametrize the momentum dependence by 
the oriented distance $q_r$ from the $2k_F$-line defined 
at fixed $\mu$ and $T=0$ as in the preceding section.
We denote the shift of the $2k_F$-line near $\bq_{2k_F}$
at $T=0$ corresponding to a chemical potential $\mu' \neq \mu$
by $q_r^0(\mu')$.
Eq.~(\ref{near2k_F}) then yields 
\begin{eqnarray} \label{Pimu'}
 \Pi_d^0(\bq;0,\mu') &=&  \Pi_d^0(\bq'_{2k_F};0,\mu') + 
 b_F [q_r - q_r^0(\mu')] \nonumber\\&&
 +a_F \Theta[q_r - q_r^0(\mu')] \sqrt{q_r - q_r^0(\mu')} \; ,
\end{eqnarray}
where $a_F = d_{\bk_F}^2 m_F^{1/2}/(2\pi \, v_F^{1/2})$.
Neglecting the weak and regular $\mu$-dependence of the height
of the ridge one can approximate $\Pi_d^0(\bq'_{2k_F};0,\mu')$
on the right hand side by $\Pi_d^0(\bq_{2k_F};0,\mu)$.
Inserting Eq.~(\ref{Pimu'}) into Eq.~(\ref{convol}), and using
the antisymmetry of $q_r^0(\mu')$ around $\mu$, that is,\linebreak
$q_r^0(\mu + \delta\mu') = - q_r^0(\mu - \delta\mu')$ for small
$\delta\mu'$, one obtains
\begin{eqnarray} \label{PiT}
 \Pi_d^0(\bq;T,\mu) &=& \Pi_d^0(\bq_{2k_F};0,\mu) + b_F q_r \nonumber\\&&
 +a_F \int_{-\infty}^{\infty} d\mu' \, h(\mu - \mu') \,\nonumber\\&&\times
 \Theta[q_r - q_r^0(\mu')] \sqrt{q_r - q_r^0(\mu')} \; .
\end{eqnarray}
The shift of the ridge at $T > 0$ is given by the solution
$q_r^p$ of the equation $\partial_{q_r} \Pi_d^0(\bq;T,\mu) = 0$,
that is,
\begin{equation} \label{shift1}
 b_F + \frac{a_F}{2} 
 \int_{-\infty}^{\infty} d\mu' \, h(\mu - \mu') \,
 \frac{\Theta[q_r^p - q_r^0(\mu')]}{\sqrt{q_r^p - q_r^0(\mu')}}
 = 0 \; .
\end{equation}
A qualitative contemplation of Eq.~(\ref{PiT}) reveals that 
$q_r^p$ is negative.
For $\mu'$ near $\mu$, $q_r^0(\mu')$ is a monotonic function of
$\mu'$. We denote its inverse function by $\mu'(q_r)$ and
linearize 
$q_r^p - q_r^0(\mu') \approx D [\mu'(q_r^p) - \mu']$, where
$D = \tfrac{\partial q_r^0}{\partial\mu'}|_{\mu' = \mu}$.
In case that $q_r^0(\mu')$ increases with $\mu'$, Eq.~(\ref{shift1})
can then be written as
\begin{equation} \label{shift2}
 b_F + \frac{a_F}{2 D^{1/2}} 
 \int_{-\infty}^{\mu'(q_r^p)} d\mu' \,
 \frac{h(\mu - \mu')}{\sqrt{\mu'(q_r^p) - \mu'}} = 0 \; .
\end{equation}
Introducing the variable $\delta\mu' = \mu' - \mu$, and substituting
$\delta\mu' = \delta\mu'(q_r^p) u$, where $\delta\mu'(q_r^p) < 0$, 
the integral in Eq.~(\ref{shift2}) can be written as
\begin{align} \label{int1}
 &\int_{-\infty}^{\mu'(q_r^p)} d\mu' \,
 \frac{h(\mu - \mu')}{\sqrt{\mu'(q_r^p) - \mu'}} \nonumber\\
 &\quad = \sqrt{-\delta\mu'(q_r^p)} \int_1^{\infty} \frac {du}{\sqrt{u-1}} \,
 \frac{1}{4T \cosh^2 \frac{\delta\mu'(q_r^p) u}{2T}} \; .
\end{align}
One can see that a solution of Eq.~(\ref{shift2}) requires
$|\delta\mu'(q_r^p)| \gg T$ for small $T$.
Therefore, we can approximate
$\cosh^2 \tfrac{\delta\mu'(q_r^p) u}{2T} \approx 
 \frac{1}{4} e^{|\delta\mu'(q_r^p)| \, u/T}$.
The remaining integral is elementary and yields
\begin{equation} \label{int2}
 \int_{-\infty}^{\mu'(q_r^p)} d\mu' \,
 \frac{h(\mu - \mu')}{\sqrt{\mu'(q_r^p) - \mu'}} =
 \sqrt{\frac{\pi}{T}} e^{-|\delta\mu'(q_r^p)|/T} \; .
\end{equation}
Inserting this in Eq.~(\ref{shift2}) yields
\begin{equation} \label{deltamu'}
 |\delta\mu'(q_r^p)| = 
 \frac{1}{2} \, T \ln \frac{T_0}{T} \; .
\end{equation}
where $T_0 = \pi a_F^2/(4 b_F^2 D)$. \pagebreak
Note that indeed $|\delta\mu'(q_r^p)| \gg T$ for small $T$, 
that is, for $T \ll T_0$.
For the shift $q_r^p$ one thus obtains
\begin{equation}
 q_r^p = - \frac{D}{2} \, T \ln \frac{T_0}{T} \; .
\end{equation}
Replacing $D$ by $|D|$, the last two equations are valid 
also in the case where $q_r^0(\mu')$ decreases with $\mu'$.
In summary, the ridge is shifted toward the inner side of the
$2k_F$-line by an amount of order $T |\log T|$.
Hence, the peaks of $\Pi_d^0(\bq,0)$ at $\bq^* = (\pm q_a,0)$ 
and $(0,\pm q_a)$ below Van Hove filling, and at 
$\bq^* = (\pm q_d,\pm q_d)$ above Van Hove filling, are also 
subject to a shift of order $T |\log T|$.

To quantify the smoothing of the peak at the $2k_F$-line at
finite temperature, we evaluate $\partial_{q_r}^2 \Pi_d^0$
at $q_r^p$.
Substituting $q_r - q_r^0(\mu') = D[\mu'(q_r) - \mu']$ and
performing a partial integration, the second derivative of
$\Pi_d^0$ with respect to $q_r$ can be written as
\begin{equation}
 \partial_{q_r}^2 \Pi_d^0 = \frac{a_F}{2D^{3/2}}
 \int_{-\infty}^{\infty} d\mu' \, h'(\mu'-\mu) \,
 \frac{\Theta[\mu'(q_r) - \mu']}{\sqrt{\mu'(q_r) - \mu'}}
 \; .
\end{equation}
For large $|\delta\mu'(q_r)|/T$ one can approximate
$h'(\mu'-\mu) \approx \sgn(D) \, T^{-2} e^{|\delta\mu'|/T}$ 
and perform the integral explicitly, to obtain
\begin{equation}
 \partial_{q_r}^2 \Pi_d^0 = 
 \frac{a_F \sqrt{\pi}}{2|D|^{3/2} T^{3/2}} \,
 e^{-|\delta\mu'(q_r)|/T} \; .
\end{equation}
Inserting $\delta\mu'(q_r^p)$ from Eq.~(\ref{deltamu'}), one
obtains the curvature at the shifted peak position
\begin{equation}
 \left . \partial_{q_r}^2 \Pi_d^0 \right|_{q_r = q_r^p} = 
 \frac{a_F \sqrt{\pi}}{2|D|^{3/2} \sqrt{T_0}} \, \frac{1}{T} =
 \frac{|b_F|}{|D|} \, \frac{1}{T} \; .
\end{equation}
The last two equations are valid for any sign of $D$.
Hence, the radius of curvature of the peak is proportional
to $T$ at small temperatures, with a remarkably simple 
prefactor.


\section{Conclusion}

We have analyzed the strength of nematic fluctuations with a
finite wave vector in a two-dimensional metal. To this end we 
have computed the bare static $d$-wave polarization function
$\Pi_d^0(\bq,0)$ as a function of the wave vector $\bq$ for 
electrons with a tight-binding dispersion on a square lattice.
Peaks in $\Pi_d^0(\bq,0)$ indicate at which wave vectors a 
(modulated) nematic instability occurs in presence of a 
sufficiently strong attraction in the $d$-wave charge channel.

At Van Hove filling, $\Pi_d^0(\bq,0)$ is strongly peaked at
$\bq = 0$, so that the leading nematic instability is
homogeneous in this case.
Below and close to Van Hove filling, the largest peaks are
on the $q_x$- and $q_y$-axes, leading to a modulated nematic
state with a small modulation vector along one of the crystal 
axes.
Above Van Hove filling, the largest peaks of $\Pi_d^0(\bq,0)$
are situated at diagonal wave vectors $\bq^* = (\pm q_d,\pm q_d)$, 
so that the dominant instability leads to a spatially modulated 
nematic state with a diagonal modulation vector.
The same modulated nematic state has been found by Metlitski 
and Sachdev \cite{metlitski10a,metlitski10b} in a recent study 
of secondary instabilities generated by antiferromagnetic
fluctuations in a two-dimensional metal. In that context the 
wave vector $\bq^*$ is favored because it connects intersections 
of the Fermi surface with the antiferromagnetic Brillouin zone 
boundary (hot spots).
Remarkably, the peak at the same $\bq^*$ in the $d$-wave 
polarization function is determined purely by the Fermi surface 
geometry, without any influence from antiferromagnetic 
fluctuations.

In all cases the peaks of $\Pi_d^0(\bq,0)$ lie on lines defined
by the condition $\eps_{(\bq + \bG)/2} = \mu$, where $\bG$ is
a reciprocal lattice vector, which is the lattice analogue of
the condition $|\bq| = 2k_F$ in a continuum system.
Generically, the momentum dependence of $\Pi_d^0(\bq,0)$ 
exhibits a square root singularity at these lines, which 
therefore characterizes also the behavior around the peaks of 
$\Pi_d^0(\bq,0)$.
At low finite temperatures the peaks in the polarazation function
are smoothed and shifted by an amount of order $T |\log T|$.

In view of the above results it seems worthwhile to search
for modulated nematic instabilities in two-dimensional 
Hubbard-type models.
In a recent functional renormalization group study of the 
one-band Hubbard model, a modulated nematic instability was 
found to be typically favorable compared to a homogeneous one, 
but in any case weaker than antiferromagnetism or $d$-wave 
superconductivity.
Adding a nearest-neighbor repulsion strengthens the nematic
fluctuations.\cite{husemann12}
There is more room for nematic instabilities in multi-band
systems.\cite{fischer11} A search for modulated nematic states 
in such systems would therefore be particularly promising.


\begin{acknowledgments}
We are very grateful to C.~Husemann, M.~Metlitski, and 
H.~Yamase for valuable discussions.
\end{acknowledgments}


\end{document}